\documentclass[letterpaper, 10 pt, conference]{ieeeconf}
\UseRawInputEncoding
\pdfoutput=1
\IEEEoverridecommandlockouts
\usepackage{array}
\usepackage[caption=false,font=normalsize,labelfont=sf,textfont=sf]{subfig}
\usepackage{textcomp}
\usepackage{stfloats}
\usepackage{url}
\usepackage{verbatim}
\usepackage{xcolor}
\usepackage{multirow}
\usepackage{tabularx}
\def\BibTeX{{\rm B\kern-.05em{\sc i\kern-.025em b}\kern-.08em
    T\kern-.1667em\lower.7ex\hbox{E}\kern-.125emX}}
\usepackage{balance}

\usepackage{algorithm}
\usepackage{algorithmic}

\makeatletter
\newcommand\fs@norules{\def\@fs@cfont{\bfseries}\let\@fs@capt\floatc@ruled
  \def\@fs@pre{}%
  \def\@fs@post{}%
  \def\@fs@mid{\kern3pt}%
  \let\@fs@iftopcapt\iftrue}
\makeatother
\floatstyle{norules}
\restylefloat{algorithm}

\usepackage{graphics} 
\usepackage{epsfig} 
\usepackage{mathptmx} 
\usepackage{times} 
\usepackage{amsmath} 
\usepackage{amssymb}  
\usepackage{amsfonts,amssymb}
\usepackage{bm}
\usepackage{graphicx}
\usepackage{float}
\usepackage{subfloat}
\usepackage[outdir=./]{epstopdf}
\usepackage{booktabs}
\usepackage{mathrsfs}

\title{\LARGE \bf
Muscle Activation Estimation by Optimizing the Musculoskeletal Model for Personalized Strength and Conditioning Training
}

\author{Xi Wu$^{1}$, Chenzui Li$^{1}$, Kehan Zou$^{2}$, Ning Xi$^{2}$ and Fei Chen$^{1}$
\thanks{*This work is supported in part by the Research Grants Council of the
Hong Kong SAR via the Grant 24209021, 14222722, 14211723, C7100-
22GF, SRFS-2022-03-01-HKU and in part by the InnoHK of the Government of the Hong Kong SAR via the Hong Kong Sports Institute. (Corresponding author: Fei Chen.)}
\thanks{$^{1}$Xi Wu, Chenzui Li and Fei Chen are with the Department of Mechanical and Automation Engineering, T-Stone Robotics Institute, The Chinese University of Hong Kong, Hong Kong SAR, China (e-mail: {\tt\small xwu@mae.cuhk.edu.hk, czli@mae.cuhk.edu.hk, f.chen@ieee.org}).}%
\thanks{$^{2}$Kehan Zou and Ning Xi are with the Department of Industrial and Manufacturing System Engineering, The University of Hong Kong, Hong Kong SAR, China (e-mail: {\tt\small u3008454@connect.hku.hk, xining@hku.hk}).}%
}

\begin{document}

\maketitle
\thispagestyle{empty}
\pagestyle{empty}

\begin{abstract}

Musculoskeletal models are pivotal in the domains of rehabilitation and resistance training to analyze muscle conditions. However, individual variability in musculoskeletal parameters and the immeasurability of some internal biomechanical variables pose significant obstacles to accurate personalized modelling. Furthermore, muscle activation estimation can be challenging due to the inherent redundancy of the musculoskeletal system, where multiple muscles drive a single joint. This study develops a whole-body musculoskeletal model for strength and conditioning training and calibrates relevant muscle parameters with an electromyography-based optimization method. By utilizing the personalized musculoskeletal model, muscle activation can be subsequently estimated to analyze the performance of exercises. Bench press and deadlift are chosen for experimental verification to affirm the efficacy of this approach.


\end{abstract}

\section{INTRODUCTION}

Strength and conditioning training is one of the most popular forms of exercise, which can improve the overall health of individuals, such as enhancing skeletal muscle health, bone mineral density, body metabolism and mental health \cite{Schuch2016ExerciseAA}, and can also improve sports performance of athletes, such as enhancing strength, explosiveness, endurance and muscle hypertrophy \cite{Kraemer2004FundamentalsOR, Xie2022ARC}. Bench press and deadlift are essential training exercises for athletes across all sports since they involve the majority of upper and lower body muscle groups, respectively, which makes them the most effective exercises for improving upper and lower body strength. 

Understanding which muscles are activated during specific training exercises is crucial for selecting the most appropriate exercises to target desired outcomes \cite{Bourne2016ImpactOE}. By analyzing muscle activation information, researchers can quantify the specific muscle recruitment mode involved in various exercises \cite{Hug2011CanMC}. For novice exercisers, this information can help guide the learning of proper exercise techniques, reducing the risk of sporting injuries. For athletes, muscle activation information can inform targeted training to develop specific muscle groups \cite{MartnFuentes2020ElectromyographicAI}. Coaches can also use this knowledge to design personalized training programs for their athletes.

There are various approaches to estimate muscle activation, which can be broadly categorized into measurement-based and non-measurement-based methods. Surface electromyography (EMG) is a non-invasive measurement tool that plays a key role in assessing muscle activation patterns. It can be defined as an electrophysiological recording technology used for the detection of the electric potential crossing muscle fiber membranes. However, the use of sEMG can be time-consuming and logistically challenging, particularly when examining a large number of muscles, as athletes would need to attach multiple sensors before each training session with assistance. 

\begin{figure}[]
\centering
\includegraphics[width=0.8\linewidth]{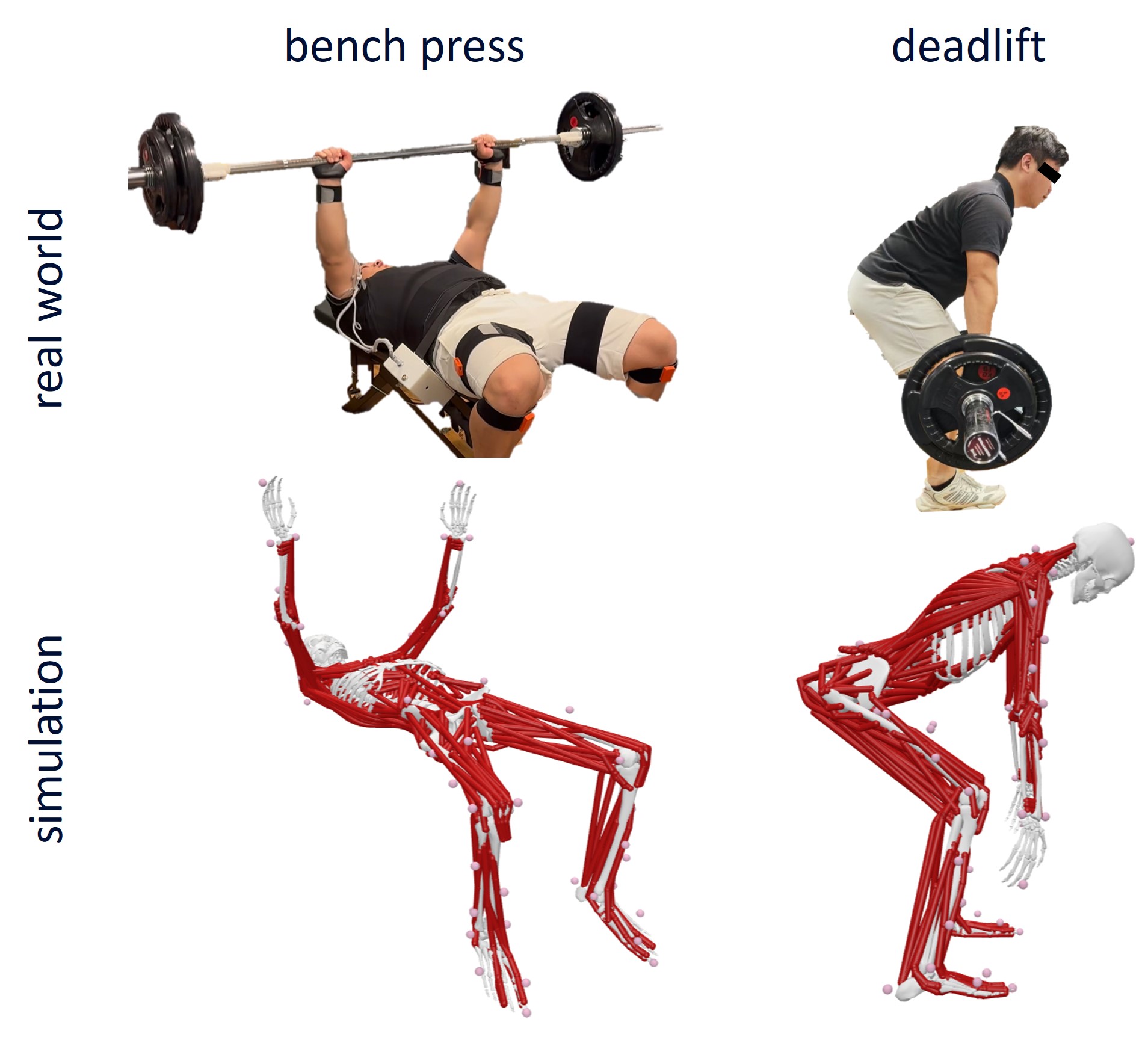} \vspace{-3mm}
\caption{Common resistance training actions and their simulation. } 
\label{Fig.resistance}\vspace{-3mm}
\end{figure} 

Considering the redundancy in the musculoskeletal system (i.e., the number of muscles exceeds the number of joints), non-measurement-based optimization methods have also been widely adopted. One of the most commonly used non-measurement approaches is static optimization, which employs a cost function based on the minimization of the sum of squared muscle activations \cite{Morrow2014ACO}. While computationally efficient, this method may not accurately represent the actual motor control strategies employed by individuals. Dynamic optimization, a optimization method aims to minimized metabolic energy and produced movement that close to reality, has been proved remarkably similar with static optimization \cite{Anderson2001StaticAD, Davy1987ADO}. Electromyography-driven modelling, which utilizes pre-existing musculoskeletal models and optimizes them based on EMG data, is a promising method \cite{SShourijeh2019MuscleSM}. However, many of these models have not been adequately personalized to the specific individuals being studied.

While muscle models have been utilized to compute muscle activation information, it is important to note that individual differences in body parameters lead to variations in personalized muscle model parameters. Pengchen Lian el. has employed exoskeleton-based measurements to optimize the parameters of a subject-specific musculoskeletal model in order to estimate the active torque at the knee joint \cite{Lian2022ATH}. Current approaches to musculoskeletal model personalization often focus on parameters such as muscle lengths and force-velocity curves, typically for single-joint, single-action scenarios without considering changes in external forces \cite{Sartori2012EMGDrivenFE}. This limited scope may not adequately capture the complex and individualized nature of human movement and muscle activation patterns.

In this paper, we propose a novel framework for estimating muscle activation during strength and conditioning training, leveraging personalized musculoskeletal modelling. In order to obtain the subject-specific musculoskeletal model, the generalized model is re-calibrated based on the anatomical parameters obtained from measurement, and the immeasurable muscle parameters are calibrated by the proposed optimization algorithm. Then another optimization method leverages subject-specific musculoskeletal model parameters to provide an accurate representation of individual muscle activation patterns towards strength and conditioning training. The effectiveness of the proposed method is validated through experimental trials involving multiple load conditions for both bench press and deadlift, which are multi-joint exercises (Fig. \ref{Fig.resistance}).

The main contributions of this work are:
\begin{itemize} 
  \item [$\bullet$]
  A novel muscle activation estimation method is proposed based on personalized musculoskeletal modelling. 
  \item [$\bullet$]
  A simplified calibration method is given for musculoskeletal models, making it more practical for real-world applications.
  \item [$\bullet$]
  Experimental validation of the proposed algorithm is carried out for bench presses and deadlifts under varying external forces.
\end{itemize}

\begin{figure*}[htbp]
\centering
\includegraphics[width=0.99\linewidth]{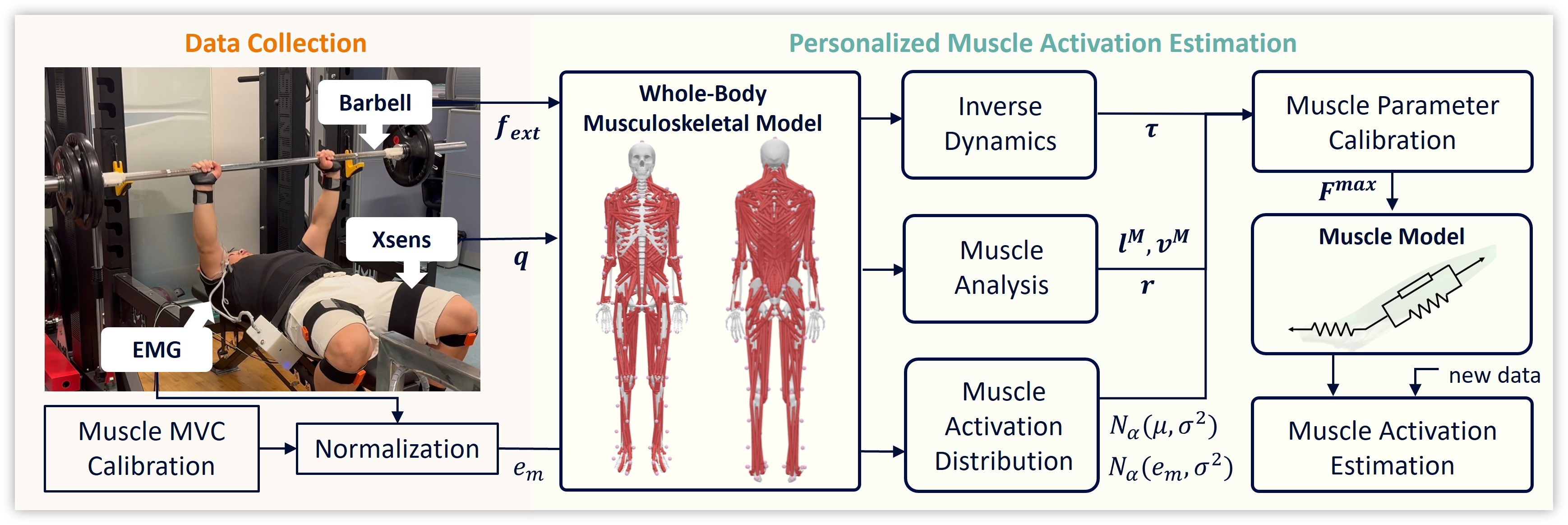} \vspace{-3mm}
\caption{The overall framework scheme. \textit{Data Collection:} Human information in strength and conditioning training is recorded by various sensors. \textit{Personalized Muscle Activation Estimation:} A whole-body musculoskeletal model is employed to perform inverse dynamics and muscle analysis using the acquired human body information. Subject-specific muscle parameter calibration is conducted in conjunction with muscle activation distributions to obtain personalized muscle models, which are then used to estimate muscle activation on new data.} 
\label{Fig.framework}\vspace{-3mm}
\end{figure*}

\section{PROPOSED METHODS}\label{methods} \vspace{-1mm}
The diagram of the overall framework is illustrated in Fig. \ref{Fig.framework}. The following sections will detail the musculoskeletal model, the method for personalizing musculoskeletal model and estimating muscle activation. 

\subsection{Musculoskeletal Model} \label{Musculoskeletal Model}

In this paper, a whole-body musculoskeletal model is build within OpenSim, which was constructed by combining the Rajagopal model \cite{rajagopal2016full} and the Bilateral Upper Extremity Trunk Model \cite{chen2023bilateral}. The model is shown as part of Fig. \ref{Fig.framework}.

To obtain a subject-specific musculoskeletal model, the generic model is scaled to match the individual’s anthropometric dimensions and proportions. Directly measured anatomical parameters, such as weight and bone lengths, are used for this scaling process. This results in a model that closely resembles the subject's size and proportions. However, certain muscle parameters cannot be obtained through direct measurement.

\begin{figure}
\centering
\includegraphics[width=0.85\linewidth]{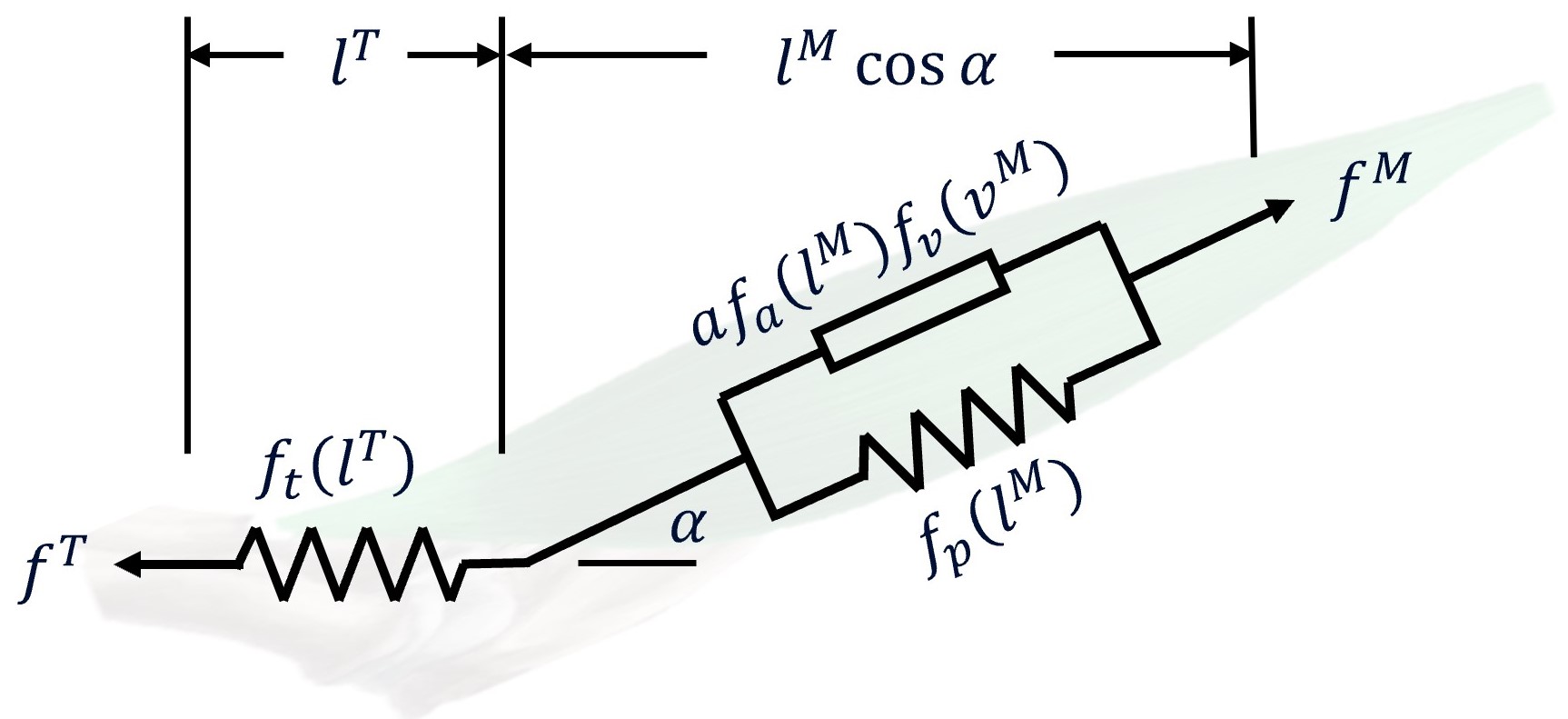} \vspace{-3mm}
\caption{Structure of Hill-type muscle model. } 
\label{Fig.muscle}\vspace{-3mm}
\end{figure}

The Hill-type muscle model, a widely adopted biomechanical approach, was utilized to represent the three-unit skeletal muscle function, consisting of an active contractile element, a passive elastic element, and an elastic tendon. The structure of Hill-type muscle model is shown in Fig. \ref{Fig.muscle}. 

If the tendon is assumed to be elastic and the mass of the muscle is assumed to be negligible, then the muscle and tendon forces must be in equilibrium:
\begin{equation}
F^{max}\left(af_a(l^M)f_v(v^M)+f_p(l^M)\right)\cos\alpha-F^{max}f_t(l^T)=0
\end{equation}

The muscle force could be written as:
\begin{equation}
F = F^{max}\left(af_a(l^M)f_v(v^M)+f_p(l^M)\right)
\end{equation}
where $a$ is the muscle activation, $F_m^{max}$ is the maximum isometric contraction force, $l^M$ is the normalized muscle length and $v^M$ is the normalized muscle velocity. $f_p(l^M)$ is the passive force–length factor, which affect the passive force of muscle. In this paper, only active force was considered, so the muscle force could be written as:
\begin{equation}
F = F^{max}a f_a(l^M) f_v(v^M)
\end{equation}
where $f_a(l^M)$ is the active force–length factor, which is close to 1 when the actual muscle length is closed to the optimal fiber length, and $f_v(v^M)$ is the force–velocity factor which is close to 1 when the muscle velocity is low. On muscle activation estimation methods, such as static optimization, previous studies have shown that the force-length-velocity properties of muscle have little influence \cite{Anderson2001StaticAD}. Therefore, the following simplified equation is commonly used in the optimization process:
\begin{equation}\label{mforce}
F = F^{max}a
\end{equation}

It could be seen that the important parameters of muscle force are the muscle activation $a$, which could be measured by EMG device, and the maximum isometric contraction force $F_m^{max}$. So maximum isometric contraction force $F_m^{max}$ will be the target to get personalized musculoskeletal model.

\subsection{Personalized musculoskeletal model}

In this subsection, individualized muscle parameters are identified by solving an optimization problem to minimize the difference between the joint torque calculated by inverse dynamics and the estimate torque based on the human model. 

The dynamics model if motion could be written as:
\begin{equation}\label{eq1}
M(q)\ddot{q}+V(q,\dot{q})+G(q)+J^T F_{ext}=\Gamma 
\end{equation}
where $q, \dot{q}, \ddot{q}$ are the vectors of joint angles, velocities, and accelerations, respectively; $M(q)$ is the system mass matrix; $V(q,\dot{q})$ is the vector of Coriolis and centrifugal forces; $G(q)$ is the vector of gravitational forces; $F_{ext}$ is the vector of external force; $J$ is the Jacobian matrix; $\Gamma = \{\tau_1, \dots ,\tau_N\}$ is the vector of joint torque.

Another method to estimate joint torque is based on the musculoskeletal model.
\begin{equation}\label{eq2}
\tau_{\mathrm{c},n}=\sum_{m=1}^{M}F_{m}r_{m,n}
\end{equation}
where $F_{m}$ is the force of $m$-th muscle; $r_{m,n}$ is the moment arm of $m$-th muscle at $n$-th joint; $\tau_{\mathrm{c},n}$ is the corresponding joint torque.

According to Eq.(\ref{mforce}), muscle activation is very important in calculating. But EMG data, the signal presents the trend of muscle activation, is usually noisy. So besides full-wave rectification, a low-pass third-order Butterworth filter at $6\ Hz$ and 
normalization applied to the raw EMG signals, distribution is also used to constrain the muscle activation. EMG data of multiple cycles is used to calculate the Gaussian distribution $N_a(\mu,\sigma^2)$ during the whole cycle, where $\mu$ is average and $\sigma$ is standard deviation. 
\begin{equation}\label{eq3}
a_m\sim N_a(\mu,\sigma^2)
\end{equation}
where $a_m$ is muscle activation of $m$-th muscle. For a specific motion cycle, processed EMG data might be more accurate, so another distribution $N_a(\mu,\sigma^2)$ is applied.
\begin{equation}\label{eq4}
a_m\sim N_a(e_m,\sigma^2)
\end{equation}

Muscle activation represents the activation level of the muscle, so the value satisfied:
\begin{equation}\label{eq5}
0\leq a_{m}\leq1
\end{equation}

Therefore, interior point method \cite{Wchter2006OnTI}, a nonlinear optimization method, could be used to solve the musculoskeletal model personalizing problem:
\begin{equation}\label{op1}
\begin{aligned}
\operatorname*{min}&\sum\left(\tau^{c}_{n}-\tau_n \right)^{2} \\
s.t.\quad&a_m\sim N_a(\mu,\sigma^2) \\
&a_m\sim N_a(e_m,\sigma^2) \\
&0\leq a_{m}\leq1 \\
&\tau^{c}_{n}=\sum_{m=1}^{M}F_{m}r_{m,n} \\
&F_m=f(F_m^{max},a_m) \\
&\alpha F_m^r\leq F_m^{max}\leq \beta F_m^r 
\end{aligned}
\end{equation}
where $F_m^r$ is the reference value of $F_m^{max}$, which is obtained from OpenSim model. The maximum isometric contraction force $F_m^{max}$ might vary greatly among different people or during different state, so a reasonable range was used to constrain $F_m^{max}$. $F_m$ is the muscle force of $m$-th muscle, which could be calculated by the mdoel in \ref{Musculoskeletal Model}.


\subsection{Muscle Activation Estimation}

After getting personalized musculoskeletal model, precise muscle activation estimation is possible. The idea used is the same as the previous subsection. Difference is the target changed and EMG device could not be used to give a reference muscle activation. As the same motion was used during personalized musculoskeletal modelling and muscle activation estimation, distribution Eq.(\ref{eq3}) could also be used.

Therefore, the muscle activation estimation problem could be summarized as:
\begin{equation}\label{op2}
\begin{aligned}
\operatorname*{min} &\sum(\tau^{\alpha}_{n}-\tau_{n})^{2} \\
s.t.\quad&\alpha_m\sim N_a(\mu,\sigma^2) \\
&0\leq a_{m}\leq1 \\
&\tau^{\alpha}_{n}=\sum_{m=1}^{M}F_{m}r_{m,n} \\
&F_m=f(F_m^{max},a_m)
\end{aligned}
\end{equation}

Eq.(\ref{op2}) is similar with Eq.(\ref{op1}), but the maximum isometric contraction force $F_m^{max}$ is known in Eq.(\ref{op2}) and the optimization target is muscle activation, while Eq.(\ref{op1}) is the opposite. 

\section{EXPERIMENT}


\begin{table*}[htbp]
\caption{Muscle selection of bench press and deadlift} \vspace{-1.5mm}
\centering
\begin{tabular}{ccm{8cm}m{5cm}}
\toprule
 & \multicolumn{1}{c}{\textbf{Joint}}  
 & \multicolumn{1}{c}{\textbf{Single-Joint Muscle}} 
 & \multicolumn{1}{c}{\textbf{Two-Joint Muscle}} \\ 
\midrule
\multirow{2}{*}[-2ex]{Bench Press} 
& Elbow 
& Brachialis (BRA), Brachioradialis (BRD), lateral head of Triceps Brachii (TriLat) 
& \multirow{2}{5cm}[-1ex]{long head and short head of Biceps Brachii (BicLong, BicSho), long head of Triceps Brachii (TriLong)} \\ \cline{2-3}
& Shoulder 
& Deltoid Anterior (DelAnt), Deltoid Medius (DelMed), Deltoid Posterior (DelPos), clavicular portion, sternal portion and costal portion of Pectoralis Major (PMCla, PMSte, PMCos), Latissimus Dorsi (LD) &
   \\ 
\hline
\multirow{3}{*}[-3ex]{Deadlift} &
  Waist &
  Erector Spinae Longissimus (ESL), Erector Spinae Iliocostalis (ESI), Multifidus (MUL), Rectus Abdominus (RA), Internal Oblique (IO), External Oblique (EO) &
  \textbackslash{} \\ \cline{2-4} 
 &
  Hip &
  Tensor Fasciae Latae (TFL), Adductor longus (AddLong), Gluteus Maximus (Gmax), Psoas Major (PM) &
  \multirow{2}{5cm}{Rectus femoris (RF), Semitendinosus (ST), Biceps femoris (BF), Gluteus Medius (Gmed)} \\ \cline{2-3}
 & Knee & Tibialis anterior (TA), Gastrocnemius Lateralis (GL), Gastrocnemius Medialis (GM), Vastus Lateralis (VL), Vastus Medialis (VM) &                  \\ 
\bottomrule
\end{tabular}
\label{table.muscle}
\end{table*}

\begin{figure*}[]
\centering
\includegraphics[width=0.95\linewidth]{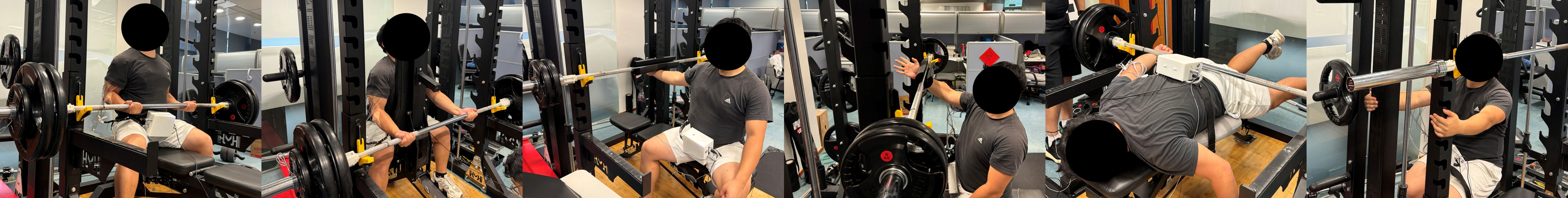}  \vspace{-1.5mm}
\caption{Illustration of the Maximum Voluntary Isometric Contraction tests.}  \vspace{-1.5mm}
\label{Fig.mvc}
\end{figure*} 

\begin{figure*}[htbp]
\centering
\includegraphics[width=0.9\linewidth]{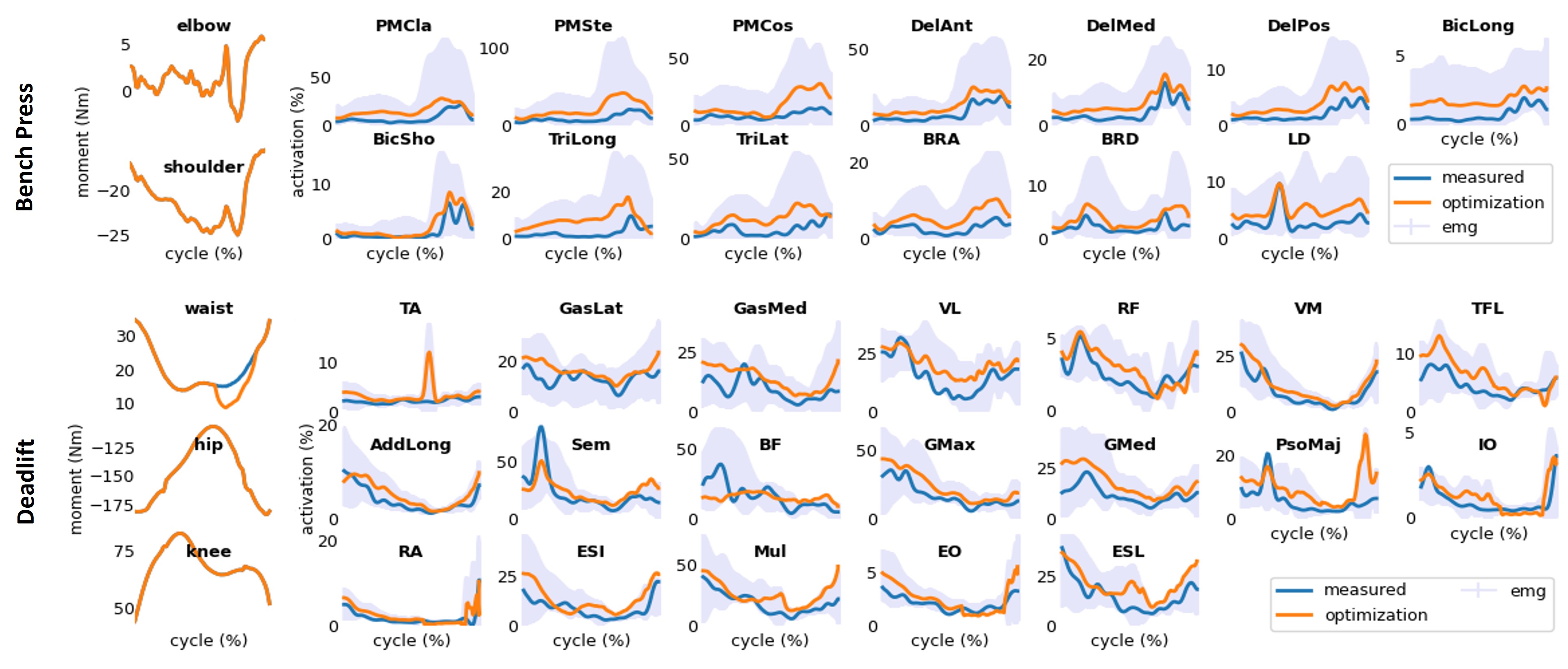}   \vspace{-3mm}
\caption{Optimization results of bench press and deadlift during musculoskeletal model personalization.}
\label{Fig.training}  \vspace{-2mm}
\end{figure*} 



Experiments were conducted on both bench press and deadlift, two fundamental and essential movements in strength training.

\subsection{Data Collection and Processing}

During the data collection process, the data of the bench press and deadlift exercises were recorded by the Xsens system and EMG sensors with the corresponding software. The motion was obtained from the Xsens system at a frequency of 60 Hz. The EMG signals were acquired through the Delsys Trigno Wireless System with a sampling rate of 2000 Hz. The placement of the EMG sensor is guided by the SENIAM ecommendations \cite{Hermens2000DevelopmentOR}. After placing the electrodes, the maximal voluntary isometric contraction (MVC) of each muscle was recorded to normalize the EMG values. The measurement method of MVC refers to \cite{Schwartz2017NormalizingSE, Fujita2022ThePM, VeraGarcia2010MVCTT}, and the actions of bench press shown in Fig. \ref{Fig.mvc}.

The weight of barbell is also recorded. The data of four load groups for bench press and deadlift were collected respectively, and each load group included 8 repetitive movements. The load groups of bench press are 20kg, 40kg, 50kg and 60kg, and the load groups of deadlift are 35kg, 45kg, 65kg and 75kg. Assumed that the barbell remains horizontal during the exercise, the force is uniformly applied to the person's hands, the left and right sides of people are symmetrical, and the force is generated in unison.

Key joints and muscles were chosen for each exercise. For the bench press, shoulder flexion/extension and elbow flexion/extension are the key joint movements, with EMG monitoring 13 muscles involved. For the deadlift, hip flexion/extension, knee flexion/extension, and waist flexion/extension are the primary joint movements, with EMG monitoring 19 muscles involved. All muscles are list in Table.\ref{table.muscle}.

\begin{table*}[]
\caption{Root Mean Square Error (RMSE) between the estimated muscle activations and the measured values in the testing set.} \vspace{-3mm}
\begin{tabular}{ccccccccccccccccccccc}
\toprule
\multicolumn{1}{c|}{} & \multicolumn{4}{c|}{bench press-20kg} & \multicolumn{4}{c|}{bench press-40kg} & 
\multicolumn{4}{c|}{bench press-60kg} & 
\multicolumn{4}{c}{bench press-50kg} \\
\multicolumn{1}{c|}{proposed} & 0.10 & 0.12 & 0.12 & 
\multicolumn{1}{c|}{0.11}     & \textbf{0.10} & \textbf{0.10} & \textbf{0.11} & 
\multicolumn{1}{c|}{\textbf{0.11}}     & \textbf{0.10} & \textbf{0.07} & \textbf{0.09} & 
\multicolumn{1}{c|}{\textbf{0.09}}     & \textbf{0.09} & \textbf{0.08} & \textbf{0.09} & \textbf{0.08} \\
\multicolumn{1}{c|}{proposed+SO} & 0.06 & 0.07 & 0.07 &  
\multicolumn{1}{c|}{0.08}        & 0.13 & 0.12 & 0.14 & 
\multicolumn{1}{c|}{0.13}        & 0.20 & 0.18 & 0.19 & 
\multicolumn{1}{c|}{0.20}        & 0.15 & 0.15 & 0.20 & 0.16 \\
\multicolumn{1}{c|}{SO}   & \textbf{0.06} & \textbf{0.07} & \textbf{0.07} &  
\multicolumn{1}{c|}{\textbf{0.08}} & 0.13 & 0.12 & 0.14 & 
\multicolumn{1}{c|}{0.13} & 0.20 & 0.18 & 0.19 & 
\multicolumn{1}{c|}{0.20} & 0.15 & 0.16 & 0.20 & 0.16 \\ 
\midrule
\multicolumn{1}{c|}{} & \multicolumn{4}{c|}{deadlift-35kg} & 
\multicolumn{4}{c|}{deadlift-45kg} & \multicolumn{4}{c|}{deadlift-75kg} & \multicolumn{4}{c}{deadlift-65kg} \\
\multicolumn{1}{c|}{proposed}    & \textbf{0.08} & \textbf{0.08} & \textbf{0.08} &  
\multicolumn{1}{c|}{\textbf{0.07}}        & \textbf{0.08} & \textbf{0.08} & \textbf{0.08} & 
\multicolumn{1}{c|}{\textbf{0.08}}        & \textbf{0.09} & \textbf{0.08} & \textbf{0.06} & 
\multicolumn{1}{c|}{\textbf{0.07}}        & \textbf{0.07} & \textbf{0.06} & \textbf{0.07} & \textbf{0.06} \\
\multicolumn{1}{c|}{proposed+SO} & 0.09 & 0.09 & 0.10 &  
\multicolumn{1}{c|}{0.09}        & 0.11 & 0.12 & 0.10 & 
\multicolumn{1}{c|}{0.10}        & 0.12 & 0.14 & 0.15 & 
\multicolumn{1}{c|}{0.14}        & 0.11 & 0.12 & 0.12 & 0.13 \\
\multicolumn{1}{c|}{SO}          & 0.12 & 0.13 & 0.14 &  
\multicolumn{1}{c|}{0.13}        & 0.15 & 0.16 & 0.14 & 
\multicolumn{1}{c|}{0.14}        & 0.19 & 0.20 & 0.21 & 
\multicolumn{1}{c|}{0.20}        & 0.17 & 0.18 & 0.17 & 0.19 \\ 
\bottomrule
 \vspace{-3mm}
\end{tabular}
\label{table.rmse}
\end{table*}

\begin{figure*}[htbp]
\centering
 \vspace{-2mm}
\includegraphics[width=0.85\linewidth]{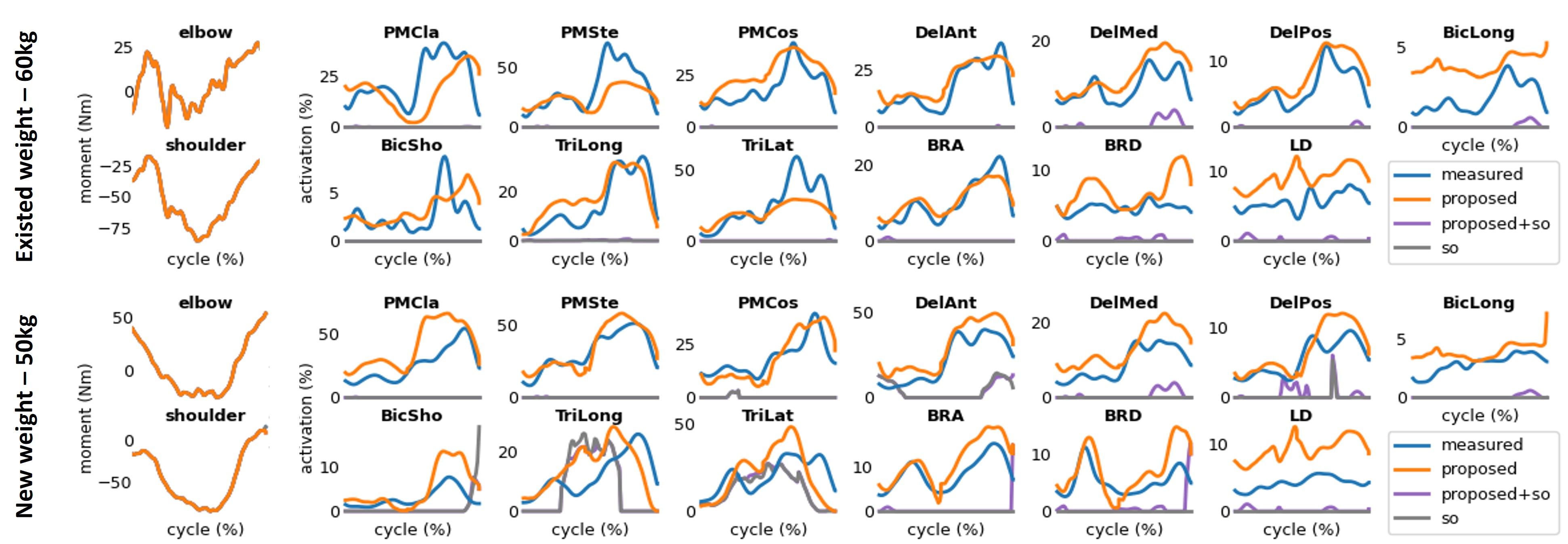}   \vspace{-2mm}
\caption{Muscle activation estimation results of bench press.} 
\label{Fig.bp-testing}\vspace{-1mm}
\end{figure*} 

\begin{figure*}[htbp]
\centering
\includegraphics[width=0.85\linewidth]{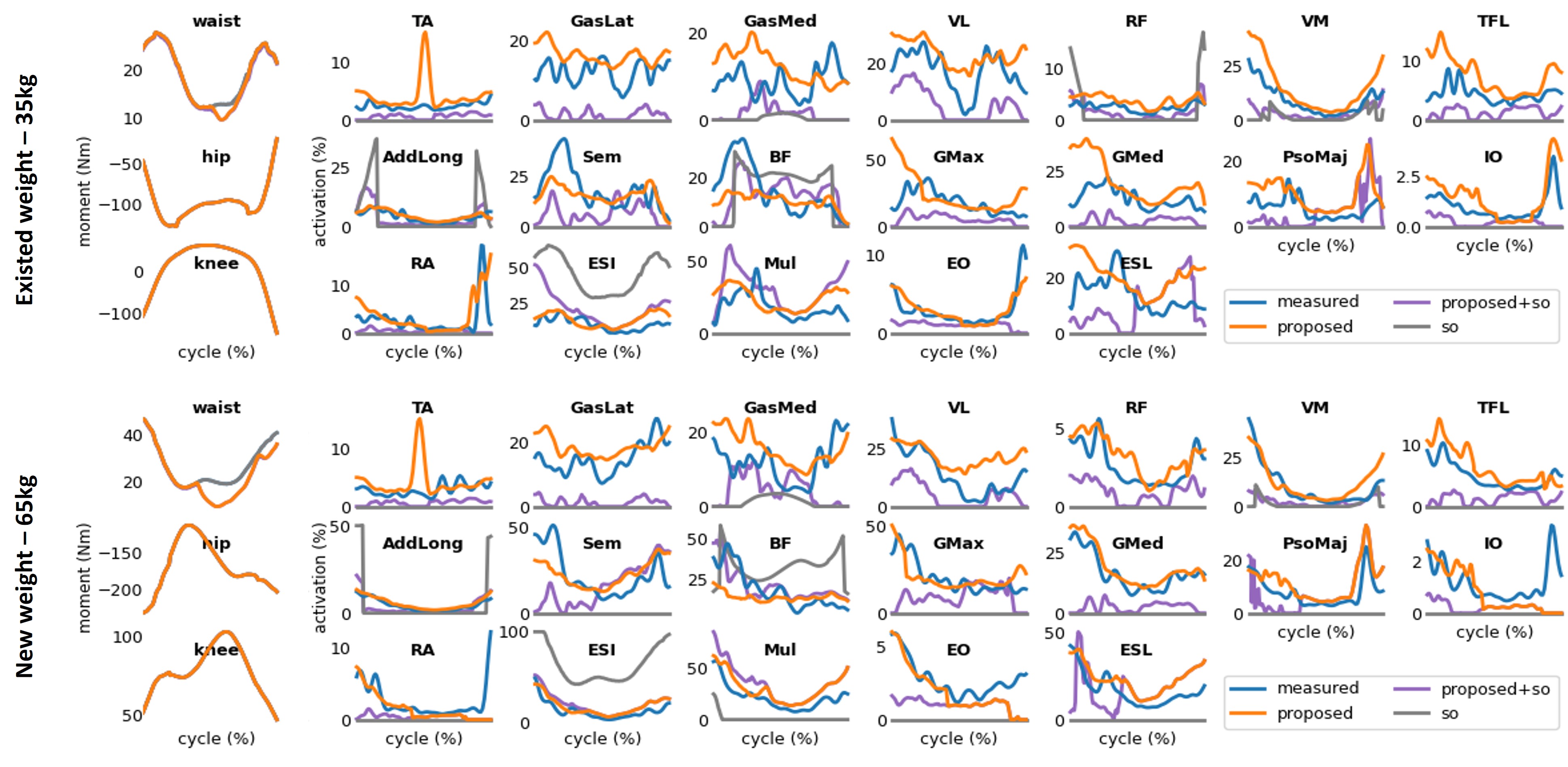}  \vspace{-3mm}
\caption{Muscle activation estimation results of deadlift.}
\label{Fig.dl-testing}\vspace{-3mm}
\end{figure*} 

\subsection{Personalizing musculoskeletal model}

To personalize the musculoskeletal model, experimental data from two dynamic tasks were used: 1) bench press with loads of 20 kg, 40 kg, and 60 kg (4 trials per load), and 2) deadlift with loads of 35 kg, 45 kg, and 75 kg (4 trials per load). These datasets served as the training set for the model optimization process.

The optimization results are presented in Fig. \ref{Fig.training}, including the optimization outcomes for a 20 kg bench press trial and a 35 kg deadlift trial. The joint moment plots, presented in the leftmost column of the figures, demonstrate a close agreement between the joint moments computed by inverse dynamics (blue line) and the optimized joint moments (orange line). Similarly, the muscle activation plots show a good correspondence between the EMG-derived muscle activations (blue line) and the optimized muscle activations (orange line), with the EMG-based muscle activation distribution also presented (light purple area). Despite the inherent noise and various factors influencing the EMG measurements, the overall trends between the optimized and measured muscle activations are consistent, indicating that the optimization process effectively captured the muscle activation patterns while maintaining accurate joint moment representations.

\subsection{Muscle Activation Estimation}

In order to validate the proposed muscle activation estimation method, we use four sets of data from four different load conditions for both the bench press and deadlift exercises as a testing set, with no overlap with the training data, to compare the measured values with the optimized values obtained using the proposed algorithm. Static optimization (SO) is a widely used method of muscle activation estimation which aims to minimize the sum of squared muscle activation. In this section, not only static optimization compared to the proposed method, but a hybrid method combining the proposed algorithm and static optimization is investigated (referred to as “proposed + SO”) was also investigated, where the objective function in Eq. (\ref{op1}) is modified to
\begin{equation}
\operatorname*{min} \left( \sum\left(\tau^{c}_{n}-\tau_{n} \right)^{2} + 0.1 \sum a^{2} \right)
\end{equation}

Fig. \ref{Fig.bp-testing} and Fig. \ref{Fig.dl-testing} present the experimental results of various methods for the bench press and deadlift exercises, respectively. Due to space constraints, only the results for one load condition from the training set and one from the testing set are shown for each exercise. Root Mean Square Error (RMSE) is used to quantify the difference between the estimated muscle activations and the measured values in the test set, with the complete data summarized in Table \ref{table.rmse}.

The analysis of Fig. \ref{Fig.bp-testing}, Fig. \ref{Fig.dl-testing}, and Table \ref{table.rmse} reveals that the estimates obtained by the proposed method are the closest to the measured values. In contrast, other methods exhibit larger errors. This suggests that the principle of minimizing the sum of squared muscle activations does not align well with the actual operating principles of the human muscle system. For the relatively heavier load conditions, the error of the static optimization increased, while the error of the proposed algorithm remained relatively small and stable. These results demonstrate the ability of the proposed method to achieve more accurate muscle activation prediction, particularly in different load conditions.

\section{CONCLUSIONS}

In this study, an optimization-based approach was developed to personalize human musculoskeletal models and estimate muscle activations during strength training exercises, such as the bench press and deadlift. By scaling the anatomical parameters of the musculoskeletal model using measured data and calibrating the difficult-to-measure muscle parameters using the proposed optimization method, the subject-specific musculoskeletal models could be constructed. These models were then used as prior knowledge in another optimization process to estimate muscle activations. Experiments were conducted on the bench press and deadlift exercises under different external load conditions, and the results validated the proposed method.

Future research will incorporate diverse muscle recruitment patterns into the personalized musculoskeletal models and expanding the application of the proposed approach to a wider range of scenarios. Additionally, based on the estimation of human’s responses to external stimuli, the constructed human musculoskeletal models can be utilized to guide the design and development of more user-centric human-machine interaction systems in the fields like ergonomics.

\addtolength{\textheight}{-12cm}   





\bibliographystyle{ieeetr}
\bibliography{ref_robio}

\begin{thebibliography}{10}

\bibitem{Schuch2016ExerciseAA}
F.~B. Schuch, D.~Vancampfort, J.~Richards, S.~Rosenbaum, P.~B. Ward, and B.~Stubbs, ``Exercise as a treatment for depression: A meta-analysis adjusting for publication bias,'' {\em physioscience}, vol.~12, pp.~122 -- 123, 2016.

\bibitem{Kraemer2004FundamentalsOR}
W.~J. Kraemer and N.~A. Ratamess, ``Fundamentals of resistance training: progression and exercise prescription.,'' {\em Medicine and science in sports and exercise}, vol.~36 4, pp.~674--88, 2004.

\bibitem{Xie2022ARC}
Z.~Xie, S.~Wang, W.~Zhao, and Z.~Guo, ``A robust context attention network for human hand detection,'' {\em Expert Syst. Appl.}, vol.~208, p.~118132, 2022.

\bibitem{Bourne2016ImpactOE}
M.~N. Bourne, M.~D. Williams, D.~A. Opar, A.~A. Najjar, G.~K. Kerr, and A.~J. Shield, ``Impact of exercise selection on hamstring muscle activation,'' {\em British Journal of Sports Medicine}, vol.~51, pp.~1021 -- 1028, 2016.

\bibitem{Hug2011CanMC}
F.~Hug, ``Can muscle coordination be precisely studied by surface electromyography?,'' {\em Journal of electromyography and kinesiology : official journal of the International Society of Electrophysiological Kinesiology}, vol.~21 1, pp.~1--12, 2011.

\bibitem{MartnFuentes2020ElectromyographicAI}
I.~Mart{\'i}n-Fuentes, J.~M. Oliva-Lozano, and J.~M. Muyor, ``Electromyographic activity in deadlift exercise and its variants. a systematic review,'' {\em PLoS ONE}, vol.~15, 2020.

\bibitem{Morrow2014ACO}
M.~M.~B. Morrow, J.~W. Rankin, R.~R. Neptune, and K.~R. Kaufman, ``A comparison of static and dynamic optimization muscle force predictions during wheelchair propulsion.,'' {\em Journal of biomechanics}, vol.~47 14, pp.~3459--65, 2014.

\bibitem{Anderson2001StaticAD}
F.~C. Anderson and M.~G. Pandy, ``Static and dynamic optimization solutions for gait are practically equivalent.,'' {\em Journal of biomechanics}, vol.~34 2, pp.~153--61, 2001.

\bibitem{Davy1987ADO}
D.~T. Davy and M.~L. Audu, ``A dynamic optimization technique for predicting muscle forces in the swing phase of gait.,'' {\em Journal of biomechanics}, vol.~20 2, pp.~187--201, 1987.

\bibitem{SShourijeh2019MuscleSM}
M.~S. Shourijeh and B.~J. Fregly, ``Muscle synergies modify static optimization estimates of joint stiffness during walking.,'' {\em Journal of biomechanical engineering}, 2019.

\bibitem{Lian2022ATH}
P.~Lian, Y.~Ma, L.~Zheng, Y.~Xiao, and X.~Wu, ``A three-step hill neuromusculoskeletal model parameter identification method based on exoskeleton robot,'' {\em Journal of Intelligent \& Robotic Systems}, vol.~104, 2022.

\bibitem{Sartori2012EMGDrivenFE}
M.~Sartori, M.~Reggiani, D.~Farina, and D.~G. Lloyd, ``Emg-driven forward-dynamic estimation of muscle force and joint moment about multiple degrees of freedom in the human lower extremity,'' {\em PLoS ONE}, vol.~7, 2012.

\bibitem{rajagopal2016full}
A.~Rajagopal, C.~L. Dembia, M.~S. DeMers, D.~D. Delp, J.~L. Hicks, and S.~L. Delp, ``Full-body musculoskeletal model for muscle-driven simulation of human gait,'' {\em IEEE transactions on biomedical engineering}, vol.~63, no.~10, pp.~2068--2079, 2016.

\bibitem{chen2023bilateral}
X.~Chen, Y.~Huang, L.~Jiang, Q.~Sun, Y.~Tian, Z.~Zhou, J.~Yin, Y.~Gao, C.~Liu, and B.~Huo, ``Bilateral upper extremity trunk model for cross-country sit-skiing double poling propulsion: model development and validation,'' {\em Medical \& Biological Engineering \& Computing}, vol.~61, no.~2, pp.~445--455, 2023.

\bibitem{Wchter2006OnTI}
A.~W{\"a}chter and L.~T. Biegler, ``On the implementation of an interior-point filter line-search algorithm for large-scale nonlinear programming,'' {\em Mathematical Programming}, vol.~106, pp.~25--57, 2006.

\bibitem{Hermens2000DevelopmentOR}
H.~J. Hermens, B.~Freriks, C.~Disselhorst-Klug, and G.~Rau, ``Development of recommendations for semg sensors and sensor placement procedures.,'' {\em Journal of electromyography and kinesiology : official journal of the International Society of Electrophysiological Kinesiology}, vol.~10 5, pp.~361--74, 2000.

\bibitem{Schwartz2017NormalizingSE}
C.~Schwartz, F.~Tubez, F.~Wang, J.-L. Croisier, O.~Br{\"u}ls, V.~Deno{\"e}l, and B.~Forthomme, ``Normalizing shoulder emg: An optimal set of maximum isometric voluntary contraction tests considering reproducibility.,'' {\em Journal of electromyography and kinesiology : official journal of the International Society of Electrophysiological Kinesiology}, vol.~37, pp.~1--8, 2017.

\bibitem{Fujita2022ThePM}
R.~A. Fujita, N.~R.~S. Silva, B.~L.~S. Bedo, and M.~M. Gomes, ``The pre-exhaustion method does not increase muscle activity in target muscle during strength training in untrained individuals,'' {\em Journal of Human Kinetics}, vol.~82, pp.~17 -- 26, 2022.

\bibitem{VeraGarcia2010MVCTT}
F.~J. Vera-Garcia, J.~M. Moreside, and S.~M. McGill, ``Mvc techniques to normalize trunk muscle emg in healthy women.,'' {\em Journal of electromyography and kinesiology : official journal of the International Society of Electrophysiological Kinesiology}, vol.~20 1, pp.~10--6, 2010.

\end{thebibliography}

\end{document}